\documentclass[%
 reprint, 
superscriptaddress,
 amsfonts, amsmath,amssymb,
 aps, 
prx,
floatfix,
longbibliography
]{revtex4-2}

\usepackage{graphicx}
\usepackage{siunitx}

\begin{document}


\title{\textbf{Sub-kelvin cryogenic wafer probing verified by on-wafer primary Coulomb blockade thermometry}}

\author{Lassi Lehtisyrjä}
\email{Contact author: lassi.lehtisyrja@vtt.fi}
\affiliation{%
 VTT Technical Research Centre of Finland Ltd, Espoo, Finland
}%

\author{Renan P. Loreto}
\altaffiliation{%
 Present address: SemiQon Oy, Espoo, Finland
}%
\affiliation{%
 VTT Technical Research Centre of Finland Ltd, Espoo, Finland
}%

\author{Juho Luomahaara}
\altaffiliation{%
 Present address: Arctic Instruments Oy, Espoo, Finland
}%
\affiliation{%
 VTT Technical Research Centre of Finland Ltd, Espoo, Finland
}%

\author{Jarno Järvinen}
\affiliation{%
 Bluefors Oy, Helsinki, Finland
}%

\author{Tuure Rantanen}
\affiliation{%
 VTT Technical Research Centre of Finland Ltd, Espoo, Finland
}%

\author{Juha Vikstedt}
\affiliation{%
AEM Afore Oy, Lieto, Finland
}%

\author{Janne S. Lehtinen}
\altaffiliation{%
 Present address: SemiQon Oy, Espoo, Finland
}%
\affiliation{%
 VTT Technical Research Centre of Finland Ltd, Espoo, Finland
}%

\author{Matti Remes}
\affiliation{%
AEM Afore Oy, Lieto, Finland
}%

\author{Timo Salminen}
\affiliation{%
AEM Afore Oy, Lieto, Finland
}%

\author{Juuso Helander}
\affiliation{%
AEM Afore Oy, Lieto, Finland
}%

\author{Aki Junes}
\affiliation{%
AEM Afore Oy, Lieto, Finland
}%

\author{Pasi Aaltonen}
\affiliation{%
AEM Afore Oy, Lieto, Finland
}%

\author{Vesa Henttonen}
\affiliation{%
AEM Afore Oy, Lieto, Finland
}%

\author{Mika Prunnila}
\altaffiliation{%
 Present address: SemiQon Oy, Espoo, Finland
}%
\affiliation{%
 VTT Technical Research Centre of Finland Ltd, Espoo, Finland
}%


\date{\today}

\begin{abstract}
Scalable cryogenic wafer level testing is emerging as an essential tool for the development of quantum devices and cryogenic microelectronics, but the main motivation, the actual operational on-wafer temperature, remains largely unverified. In this work, utilizing a cryogenic wafer prober (CWP) enables wafer-scale electrical testing of Coulomb blockade thermometers (CBTs), while the CBTs simultaneously provide an independent static and dynamic assessment of the prober's thermal performance through primary and \textit{in situ} calibrated secondary electron thermometry, realizing a reciprocal characterization approach. Using recently introduced scalable normal metal TiW-based CBTs, which require no superconductivity suppressing magnetic fields, we have measured on-wafer electron temperatures of 630–660 mK across a full 150 mm wafer, representing, to our knowledge, record-low electron temperatures measured in a CWP. These exceed the nominal 600 mK chuck setpoint by approximately 40 mK on average, demonstrating the necessity of direct on-wafer thermometry. Self-calibrated secondary CBT thermometry resolves the thermal transients caused by chuck movement and reveals a thermal settling time within 5 minutes after probe touch down. This reciprocal wafer-scale approach simultaneously evaluates the device under test and the thermal performance of the measurement environment, providing a practical methodology for scalable sub-kelvin characterization of quantum devices and cryogenic microelectronics.

\end{abstract}

\maketitle
\section{\label{sec:intro}Introduction}
Scaling quantum technologies toward practical applications such as quantum computers requires integrating large numbers of quantum components on a single wafer, creating a corresponding need for scalable cryogenic electrical testing. Statistical data on device parameters are essential to achieve process control and high yield for device fabrication. However, cryogenic characterization remains a major throughput bottleneck. A typical test cycle that includes wafer dicing, chip mounting, wire bonding, and a cooldown–warm up cycle can take days per batch, and traditional cryostats can accommodate only a handful of samples simultaneously.

Cryogenic wafer probing was identified as an opportunity for mass device characterization already in the 1980s \cite{gearyCryogenicWaferProber1983}, with the need for testing in large-scale superconducting electronics becoming apparent in the 1990s \cite{abelsonManufacturabilitySuperconductorElectronics1999}. Recent advances in dry cryocoolers and the rise of quantum computing have driven the development of commercial solutions for industry-standard wafer sizes \cite{MicroXactCryogenicProbe2026, formfactorinc.IQ3000, blueforsoyCryogenicWaferProber}, which have already demonstrated value in large-scale quantum device characterization \cite{neyensProbingSingleElectrons2024, westWaferScaleCharacterizationSuperconductor2022, pillarisettyHighVolumeElectrical2019, candidoInvestigation300mmProcess2025, contaminMethodologyEfficientCharacterization2022}.

The performance of many quantum devices is strongly temperature dependent. For example, several key figures of merit of superconducting qubits depend on the thermal environment \cite{simbierowiczInherentThermalNoiseProblem2024b}. At dilution refrigerator temperatures, electron–phonon coupling is weak, and the on-chip electron temperature can significantly exceed the cryostat temperature \cite{wellstoodHotelectronEffectsMetals1994}. Accurate on-chip thermometry is therefore essential for quantifying the thermal environment experienced by quantum devices and cryogenic electronic circuits and for interpreting their behavior. For cryogenic characterization on wafer-scale, this information is just as crucial, as the true on-wafer temperature cannot be captured by just measuring the chuck temperature, which doesn't account for the wafer to chuck thermal resistance or thermal loads from radiation or probe contact.

The Coulomb blockade thermometer (CBT) is particularly attractive for cryogenic on-chip thermometry because it can operate as a primary thermometer, it is relatively simple to fabricate, it does not require external calibration, and it is insensitive to magnetic fields \cite{pekolaCoulombBlockadebasedNanothermometry1998}. CBTs have been demonstrated in a wide temperature range from below \qty{1}{\milli\kelvin} to above \qty{60}{\kelvin} \cite{meschkeAccurateCoulombBlockade2016,samaniMicrokelvinElectronicsPulsetube2022}. Despite these attractive properties, cryogenic wafer-scale characterization of CBT devices has not previously been demonstrated.

To date, CBTs have been fabricated predominantly using aluminum tunnel junctions due to their favorable properties \cite{clarkeSQUIDHandbook2004}. The transition temperature of superconductivity in aluminum is approximately \qty{1.2}{\kelvin} \cite{cochranSuperconductingTransitionAluminum1958}, so sub-kelvin aluminum-based CBTs need measures to suppress superconductivity, such as external magnetic fields, magnetic impurity doping \cite{jalkanenSuperconductivitySuppressionFeimplanted2005}, or proximity effect engineering \cite{koskiLaterallyProximizedAluminum2011a}. These complicate both their use as on-chip thermometers and their integration with magnetically sensitive technologies such as qubits, superconducting quantum interference devices (SQUIDs), single-flux-quantum (SFQ) circuits, and kinetic inductance devices (KIDs) \cite{rowerEvolution$1F$2023,braginskiSuperconductorElectronicsStatus2019}.

In this work, we present wafer-scale normal-metal CBT devices and their sub-kelvin characterization on a \qty{300}{\milli\meter} cryogenic wafer prober platform. Through cryogenic wafer level testing of CBT devices on a full \qty{150}{\milli\meter} wafer, we were able to quantify steady-state temperature offsets and transient dynamics of the CWP and demonstrate on-wafer electron temperatures of \qtyrange{630}{660}{\milli\kelvin}. To the best of our knowledge, this is the lowest on-wafer temperature demonstrated in a cryogenic wafer prober as determined directly by primary thermometry. The CBT characterization provides wafer-scale statistics of our recently introduced scalable normal metal TiW junction process \cite{luomahaaraScalableNonsuperconductingTunnel2026} in a representative application environment. Our results also constitute a reciprocal measurement configuration, where the thermal characteristics of the CWP and the electrical characteristics of the CBTs are gauged simultaneously in a single measurement.

\section{\label{sec:bg}Background}
\subsection{\label{subsec:cbt}CBT Theory}
A CBT comprises a two-dimensional array of normal-metal tunnel junctions, separated by metallic islands, with $N$ junctions connected in series in each row and $M$ rows connected in parallel. As first demonstrated by Pekola \textit{et. al} in 1994 \cite{pekolaThermometryArraysTunnel1994}, a CBT can be described as a two-junction single-electron transistor (SET) in the weak Coulomb blockade regime $E_C \ll k_\mathrm{B}T$, and modeled by the orthodox SET theory, known as the CBT master equation. 

The charging energy of the array is given as

\begin{equation}
    \label{eq:ec}
    E_\mathrm{C} = 2\frac{N-1}{N}\frac{e^2}{2C_{\Sigma}},
\end{equation}

where $e$ is the electron charge and $C_\Sigma$ is the total island capacitance. The prefactor $2\frac{N-1}{N}$ accounts for the series connection of the CBT islands while $\frac{e^2}{2C_{\Sigma}}$ is the capacitive charging energy of a single CBT island. The conductance ratio as a function of DC bias voltage $V$ is

\begin{equation}
    \label{eq:cond}
    G(V)/G_\mathrm{T}=1-u_Ng\!\left(\frac{eV}{Nk_\mathrm{B}T}\right).
\end{equation}

Here, $G_\mathrm{T}$ is the asymptotic high-bias conductance, $u_N = E_\mathrm{C}/k_\mathrm{B}T$ is a dimensionless parameter, $k_\mathrm{B}$ is the Boltzmann constant, $T$ is the electron temperature, and $g(x)=\frac{x\sinh(x)-4\sinh^2(x/2)}{8\sinh^4(x/2)}$. This yields a bell-shaped conductance dip around zero bias voltage. The full width at half minimum of the dip provides the primary thermometry relation

\begin{equation}
    \label{eq:v_half}
    V_{1/2} \simeq 5.439\, Nk_\mathrm{B}T/e,
\end{equation}

which depends only on $N$, fundamental constants, and temperature, enabling calibration-free thermometry. 

The relative zero-bias conductance dip depth $\Delta G/ G_\mathrm{T}$ is given by

\begin{equation}
    \label{eq:secondary}
    \Delta G / G_\mathrm{T} = \tfrac{1}{6}u_N,
\end{equation}

providing a secondary thermometry relation once $E_\mathrm{C}$ and $G_\mathrm{T}$ have been determined from a primary measurement.

Higher-order corrections extending the applicable range for the master equation model toward the intermediate Coulomb blockade regime ($E_\mathrm{C} \sim k_\mathrm{B}T$), i.e., lower temperatures, are given in \cite{farhangfarOneDimensionalArrays1997}. In this regime, stochastic background charges start to significantly influence the measured device conductance \cite{feshchenkoPrimaryThermometryIntermediate2013}.

CBT thermometry directly probes the electron temperature of the metallic islands. For general purpose thermometry, where the phonon bath temperature is the quantity of interest, the electron and phonon systems must be sufficiently well thermalized. The validity of this assumption for CBT devices is discussed in Appendix~\ref{app:experiment}.

\subsection{\label{subsec:cwp}Cryogenic wafer-level testing}

Wafer-level testing (WLT) enables electrical characterization of devices directly on the wafer prior to dicing and packaging, providing high-throughput, automated measurements with statistical coverage across the full wafer. Extending WLT to cryogenic temperatures is of growing interest for quantum devices, but introduces substantial challenges. The prober must maintain a stable target temperature by controlling heat loads from mechanical motion, thermal radiation, and electrical wiring. At the same time, it must suppress electromagnetic interference and provide accurate, repeatable probe contact and automated positioning.

Mechanically, probe tips must be positioned with subdegree rotational precision and micron-scale accuracy in all three spatial dimensions. Maintaining this alignment during cooldown is complicated by differential thermal contraction, necessitating active alignment and compensation routines. Temperature uniformity across the wafer is equally critical. Thermal gradients from nonuniform wafer thermal contact to the chuck and heat loads from the environment or probe contacts can introduce systematic errors in the device under test (DUT) temperature, necessitating careful design of the thermal environment, active temperature control, and on-wafer sensing.

\section{Experimental methods}
\subsection{Device realization}
\begin{figure}[hb] 
    \centering
    \includegraphics[width=\columnwidth]{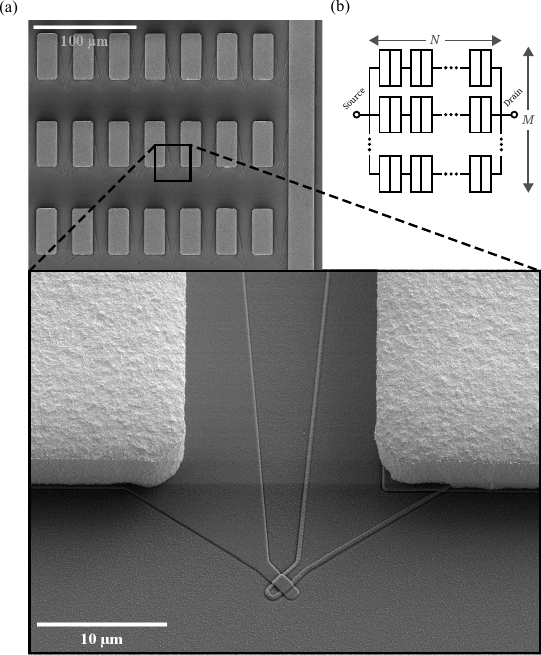}
    \caption{(a) Scanning electron micrograph of part of the CBT array. Each metallic island has nominal lateral dimensions of \qtyproduct{20 x 45}{\micro\meter}, and the copper island metallization is \qty{10}{\micro\meter} thick. The highlighted region contains a single tunnel junction between adjacent islands. (b) Circuit representation of the CBT array, comprising $N$ junctions in series and $M$ rows in parallel.}
    \label{fig:device}
\end{figure}

In this work, we employ the CBT design and TiW/Al-AlO$_\mathrm{x}$/TiW normal-metal -- insulator -- normal-metal (NIN) junction process introduced in \cite{luomahaaraScalableNonsuperconductingTunnel2026}, having a nominal junction size of \qtyproduct{0.95 x 0.95}{\micro\meter} and array dimensions of $N \times M = 61 \times 20$. A scanning electron micrograph of a part of the CBT array is presented in Fig.~\ref{fig:device}. Experiments were carried out with one wafer, containing a total of 24 reticles of size \qtyproduct{20 x 20}{\mm}, each including one CBT device equipped with a probe card compatible pad layout. Details on the device design and fabrication are elaborated in Appendix~\ref{app:design}.

To enable cryogenic wafer probing, additional processing steps were performed after the junction fabrication process. A \qty{250}{\nm} SiO$_\mathrm{x}$ passivation layer was deposited by plasma-enhanced chemical vapor deposition (PECVD), followed by via etching to expose the CBT islands. Then, \qty{20}{\nm} of TiW and \qty{300}{\nm} of copper were deposited on the islands and contact pads to act as a seed layer for subsequent electroplating of \qty{10}{\um} of copper. Only 12 of the 24 reticles on the wafer were electroplated. The copper serves a dual purpose: Its large volume and high electron-phonon coupling constant ensure good electron thermalization, while its mechanical softness enables consistent, low-resistance probe contacts. Room-temperature probing of test junctions across the wafer yielded a mean specific junction resistance of \qty{1.988}{\kohm\micro\meter\squared}.

\subsection{\label{subsec:measurements} Cryogenic wafer probing and conductance measurement}
\begin{figure}[hb] 
    \centering
    \includegraphics[width=\columnwidth]{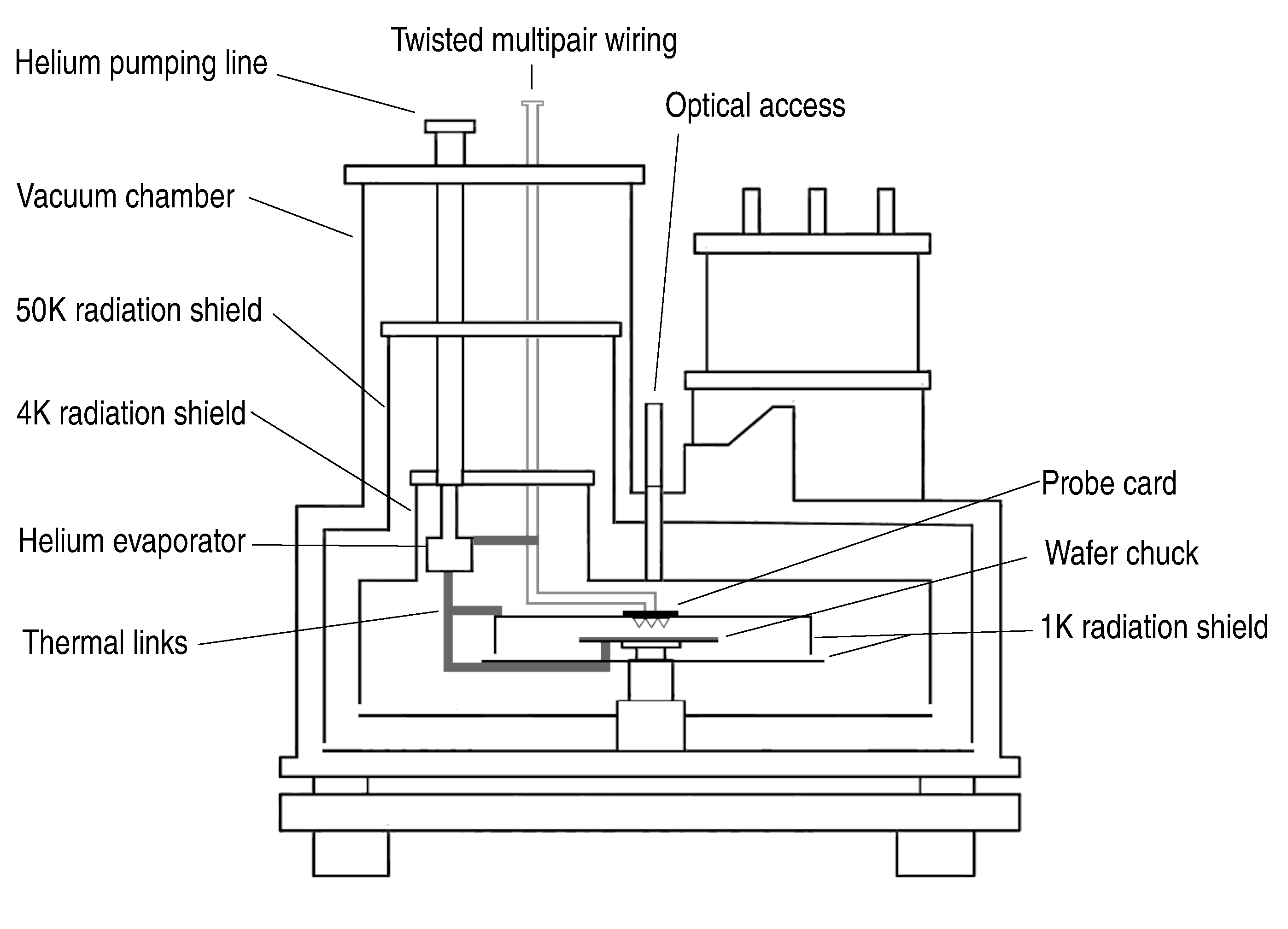}
    \caption{Schematic cross-section of the cryogenic wafer prober (CWP), illustrating the nested radiation shields and the thermal coupling of the $^3$He cooling stage to the wafer chuck and probe card. Electrical wiring and optical access for wafer alignment are also indicated.}
    \label{fig:prober}
\end{figure}

The measurements were conducted using a cryogen-free Cryogenic Wafer Prober (CWP), developed by Bluefors and AEM Afore \cite{blueforsoyCryogenicWaferProber}, which accommodates wafers up to \qty{300}{\milli\meter}. The system used in this work incorporated pre-commercial radiation-shielding and thermalization schemes, enabling a nominal chuck temperature of approximately \qty{600}{\milli\kelvin}, as measured by a calibrated resistance thermometer. Because of this shielding configuration, automated wafer transfer through the load lock was not available. A schematic cross-section of the CWP and its principal thermal and mechanical components is shown in Fig.~\ref{fig:prober}.

The CBT response of the devices was characterized by measuring the differential conductance as a function of the source-drain DC bias voltage with a standard lock-in setup in the four-terminal, voltage-biased configuration. A detailed description of the CWP, instrumentation, experiment setup, and probing procedure is presented in Appendix~\ref{app:experiment}.

\section{Results and discussion}
\begin{figure}[hb] 
    \includegraphics[width=\columnwidth]{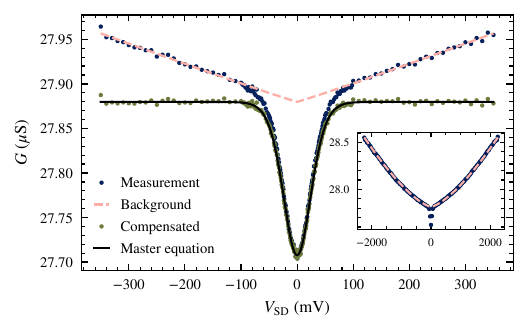}
    \caption{Differential conductance of the CBT at reticle (1,3) as a function of source-drain bias voltage at $T_\mathrm{chuck} = \qty{1.95}{\kelvin}$. The composite model of Eq.~\eqref{eq:Simmons} is fitted to the complete sweep. The dashed line shows the fitted voltage-dependent background, and division by the background factor yields the compensated conductance. The CBT master equation component of the fit is overlaid on the compensated data. The inset shows an extended sweep to $V_\mathrm{SD} = \qty{\pm2.25}{\volt}$, demonstrating the conductance nonlinearity not clearly apparent at smaller sweep amplitudes.}
    \label{fig:Gcomp}
\end{figure}

Differential conductance was measured for the 12 Cu plated CBTs over the source-drain bias range $V_\mathrm{SD} = \qty{\pm350}{\milli\volt}$ using an AC excitation voltage of \qty{1}{\milli\volt} peak-to-peak. Measurements were performed at nominal chuck temperatures of \qtylist{0.60;1.05;1.25;1.45;1.65;1.95}{\kelvin}. Measurements exhibiting clear excess noise were excluded from the corresponding analysis, and the number of devices retained therefore varies between temperature points. All acquired data, including the excluded measurements, are provided in the supporting dataset \cite{lehtisyrjaeCryogenicWaferProbingZenodo2026}. A representative CBT response measured for the device in reticle (1,3) at \qty{1.95}{\kelvin} is shown in Fig.~\ref{fig:Gcomp}, where the characteristic bell-shaped Coulomb blockade dip is clearly visible. However, the data also reveal a bias voltage-dependent background conductance, which increases in prominence at higher bias voltages and elevated temperatures. This background distorts the shape of the CBT dip and prevents reliable master equation fitting and extraction of the fitting parameters. Addressing this background distortion is required before performing any thermometric analysis.

\begin{figure*}[t!] 
    \includegraphics[width=\linewidth]{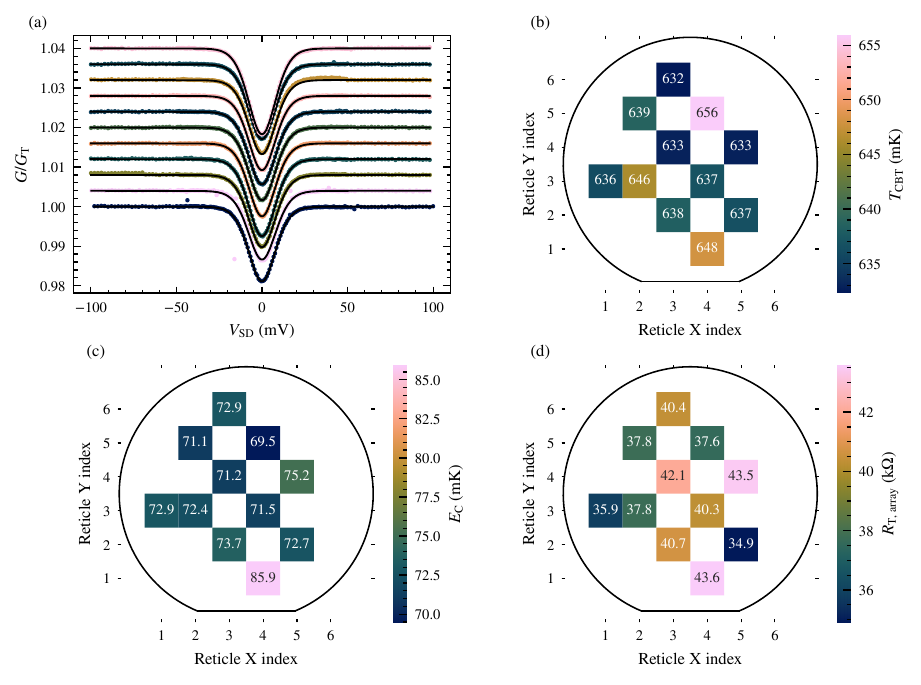}
    \caption{(a) Normalized differential conductance of the measured CBTs at $T_\mathrm{chuck} = \qty{600}{\milli\kelvin}$ after compensation for the voltage-dependent background. Solid lines show third-order CBT fits used for primary thermometry. Traces are vertically offset for clarity. Wafer maps of the fitted (b) electron temperature $T_\mathrm{CBT}$, (c) charging energy $E_\mathrm{C}$, and (d) total array resistance $R_{\mathrm{T, array}}$.}
    \label{fig:G_and_wafermaps}
\end{figure*}

\subsection{Background conductance compensation, CBT master equation fitting and wafer maps}
\label{sec:simmons}
The origins of the background conductance lie in the finite height of the junction tunnel barrier and other nonideal phenomena, which have been previously observed also in other CBTs \cite{meschkeAccurateCoulombBlockade2016, koppinenCompleteStabilizationImprovement2007, hirviOneDimensionalArrays1997}. The finite barrier is commonly modeled through the Simmons framework \cite{simmonsGeneralizedThermalJV1964}. Because the standard Simmons model did not adequately describe the measured background, we adopted an empirical composite model in which the CBT response is multiplied by a symmetric voltage-dependent background factor

\begin{equation}
\label{eq:Simmons}
    G(V) = G_{\mathrm{CBT}}(V)\,\bigl(1+\alpha\,|V-V_{\mathrm{o}}|^{\,n}\bigr).
\end{equation}

Here, $G_{\mathrm{CBT}}(V)$ is the conductance predicted by the CBT master equation Eq.~\eqref{eq:cond}, $\alpha$ is a scaling factor, $V_\mathrm{o}$ is the offset voltage (present in real measurement data), and $n$ is an exponent constrained to $0.5 \leq n \leq 2$. The constraint is physically motivated by several overlapping phenomena that give rise to a voltage-dependent background conductance: Impurity scattering at the metal-insulator interface gives a square-root dependence ($n = 0.5$) \cite{altshulerZeroBiasAnomaly1979}, inelastic tunneling describes a linear background ($n = 1$) \cite{kirtleyLinearConductanceBackgrounds1992}, and the standard Simmons model for elastic tunneling gives a quadratic dependence ($n = 2$) \cite{simmonsGeneralizedThermalJV1964}. In real devices with imperfect electrode materials and disordered oxide dielectrics, the intermediate values of $n$ reflect the coexistence of multiple mechanisms, where in addition to direct elastic tunneling, tunneling through trap-assisted hopping states and impurities in the barrier and electrode-barrier interface also may contribute \cite{bermonConductancePeaksProduced1978, glazrnanInelasticTunnelingThin1988, xuDirectedInelasticHopping1995, moranTransportPropertiesUltrathin2003}.

The composite model of Eq.~\eqref{eq:Simmons} is fitted to each measured sweep in three stages. First, the background parameters $\alpha$ and $n$ (together with $G_\mathrm{T}$)
are estimated from the data points beyond a threshold voltage $V_\mathrm{th} = \pm 2 \times V_{1/2}$, where the Coulomb blockade contribution given by the master equation (Eq.~\eqref{eq:cond}) has decayed to a negligible level. Second, the CBT parameters ($T$, $C_\Sigma$ and $V_\mathrm{o}$) are fitted with the background parameters frozen. Finally, using the parameter values obtained from the previous stages as initial values, all parameters are fitted using the composite model in Eq.~\eqref{eq:Simmons} to the measured conductance, providing the final parameter values and their uncertainties. Fig.~\ref{fig:Gcomp} illustrates a typical fitting result. The contribution of the background conductance (factor $\bigl(1+\alpha\,|V-V_{\mathrm{o}}|^{\,n}\bigr)$ in Eq.~\eqref{eq:Simmons}) is shown as a dashed line. The background-compensated conductance is then obtained by dividing the measured data by this fit, with the master-equation fit $G_\mathrm{CBT}$ overlaid.

The self-consistency of the conductance background model and fitting is supported by the stability of the extracted tunneling resistance $R_\mathrm{T, array}$, which varies by less than \SI{\pm0.1}{\percent} per device across the full  temperature range from \qtyrange{0.6}{1.95}{\kelvin} (Fig.~\ref{fig:parameters}(c)). We note that the compensation becomes less precise at elevated temperatures, where broadening of the CBT dip increases the correlation between the background and CBT parameters due to the limited number of data points outside the CBT dip range, increasing the uncertainty of the exponent $n$ in particular.

The functionality of the model was further tested on the CBT device at reticle (1,3) with a voltage sweep that extended to $V_\mathrm{SD} = \qty{\pm2.25}{\volt}$ (\qty{\sim37}{\milli\volt} per junction). There, the model gives a good fit to $n \simeq 1.4$ with the fitting result shown in the inset of Fig.~\ref{fig:Gcomp}.

We base our CBT master equation models on the Python implementation of \cite{yurttagulIndiumHighCoolingPowerNuclear2019}, adapted for our dataset. The application of the master equation for CBT parameter extraction is well justified here, as in our measurements, the devices operate firmly in the universal regime ($k_\mathrm{B}T/E_\mathrm{C} > 1$).

The compensated conductance data and the corresponding master equation fits at a nominal chuck temperature of \qty{600}{\milli\kelvin} are presented in Fig.~\ref{fig:G_and_wafermaps} (a), demonstrating an excellent fit for all devices. Further, the extracted CBT parameters of the on-wafer electron temperature $T_\mathrm{CBT}$, the charging energy $E_\mathrm{C}$, and the total tunneling resistance $R_\mathrm{T, array}$ are displayed as wafer maps in Fig. ~\ref{fig:G_and_wafermaps} (b)-(d), visualizing their spatial distribution across the \qty{150}{\mm} wafer. The temperature distribution in $T_\mathrm{CBT}$ is consistent with local temperature variations previously measured in the CWP. Such variations can arise from slight changes in the thermal radiation load as the CWP moves between DUTs and also remain within the typical uncertainty of CBT thermometry \cite{hahtelaInvestigationUncertaintyComponents2013}.  The spatial variation in $R_\mathrm{T, array}$ and $E_\mathrm{C}$  across the wafer are consistent with device-to-device variations in junction dimensions and tunnel barrier properties of the SWAPS tunnel junction process.

\subsection{Temperature dependence of the extracted parameters}
\begin{figure} 
    \includegraphics[width=\linewidth]{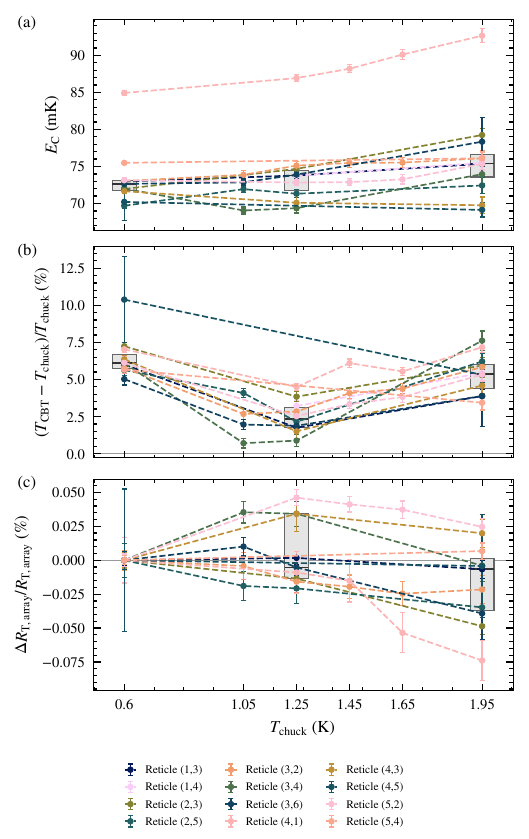}
    \caption{Temperature dependence of the extracted parameters. (a) The CBT charging energy $E_\mathrm{C}$, (b) relative temperature difference between the CBT and chuck $(T_\mathrm{CBT} - T_\mathrm{chuck})/{T_\mathrm{chuck}}$ and (c) the relative change of the tunneling resistance compared to the \qty{600}{\milli\kelvin} value $\Delta R_\mathrm{T, array}/R_\mathrm{T, array}$ as a function of chuck temperature $T_\mathrm{chuck}$ for all measured CBTs. The box plots visualize the distribution of the parameters at the temperature points with a statistically significant number of devices. The center line indicates the median, the box spans the 25th to 75th percentile distribution. The dots and dashed lines visualize individual CBT devices and their changes over temperature.}
    \label{fig:parameters}
\end{figure}

The full parameter extraction was performed across all six chuck temperature points with the results collected in Fig.~\ref{fig:fulldata} in Appendix~\ref{app:data}. The temperature dependencies of the parameters are summarized in Fig.~\ref{fig:parameters}.

The charging energy $E_\mathrm{C}$ (Fig.~\ref{fig:parameters} (a)) shows some variation with temperature across the measured devices, although the data do not conclusively establish a systematic trend across all devices. A temperature dependence in the permittivity of the amorphous AlO$_\mathrm{x}$ dielectric is one recognized mechanism that can affect junction capacitance in this regime \cite{sedehFinitebiasCoulombBlockade2025, fritzCorrelatingNanostructureAloxide2018}.

The measured on-wafer temperatures ($T_\mathrm{CBT}$) systematically exceed the chuck temperature ($T_\mathrm{chuck}$) reading. At the lowest setpoint of \qty{600}{\milli\kelvin}, the average chuck to wafer temperature offset is \qty{40}{\milli\kelvin}, and the relative chuck to wafer temperature difference remains largely constant throughout the setpoint range (Fig. \ref{fig:parameters} (b)). Possible contributions to this offset are sources of thermal load incident on the wafer, for example, from residual thermal radiation, imperfect thermal contact between the wafer and chuck, and other sources, such as uncertainty in the background conductance compensated composite model fitting and calibration or spatial offsets related to the chuck resistance thermometer. 

The total CBT array resistance $R_\mathrm{T, array}$ (Fig.~\ref{fig:parameters} (c)) varies by less than \qty{0.1}{\percent} per device over the measured temperature range. This is consistent for NIN tunnel junctions well below the energy scales for thermally activated transport. This supports the assumptions and the use of the underlying CBT master equation model. A temperature-independent $R_\mathrm{T}$ is also a prerequisite for the use of a CBT as a secondary thermometer, ensuring that the zero-bias resistance only varies based on Coulomb blockade.

\subsection{Secondary thermometry during thermalization}

\begin{figure}[h]
    \includegraphics[width=\linewidth]{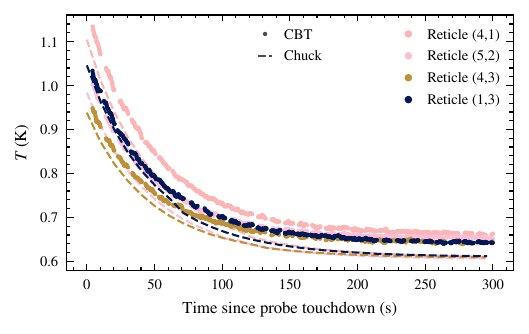}
    \caption{On-wafer CBT temperature as a function of time since probe contact for CBT devices at reticles (4,1), (5,2), (4,3), and (1,3), at a chuck temperature setpoint of \qty{600}{\milli\kelvin}. The on-wafer CBT temperature is depicted with the dots. The chuck temperature recorded by the resistance thermometer is shown as dashed lines. Each color corresponds to a single probe touchdown.}
    \label{fig:thermalization}
\end{figure}

The zero-bias conductance of each CBT device was recorded from the moment of probe contact until the start of the full differential conductance sweep, enabling secondary CBT thermometry during the thermalization period. A thermalization delay of 5 minutes was applied after each chuck movement at all temperatures, as chuck motion was observed to cause a transient temperature increase, most pronounced at the lowest temperature setpoints. Thermalization traces for four CBT devices at a chuck setpoint of \qty{600}{\milli\kelvin} are presented in Fig.~\ref{fig:thermalization}, with the measured zero-bias conductance translated to temperature via secondary CBT thermometry (Eq.~\eqref{eq:secondary}), using the parameters extracted from the master equation fits presented in Fig.~\ref{fig:G_and_wafermaps}. 

The results validate the use of a 5 min thermalization period prior to the primary thermometry measurement. Furthermore, the on-wafer CBT temperature follows the chuck temperature with a similar time constant, indicating good thermal contact between the wafer and the chuck. After thermalization, the CBT secondary temperatures are in good agreement with the primary thermometry results, despite the slow residual cooling of the chuck during the primary temperature sweep.

\section{Conclusions}
We have demonstrated a reciprocal wafer-scale characterization scheme in which a cryogenic wafer prober (CPW) performs sub-kelvin electrical testing of Coulomb blockade thermometers (CBT) across a full \qty{150}{\mm} wafer, while the CBTs simultaneously allow quantifying the thermal performance of the CPW itself. We measured on-wafer electron temperatures of \qtyrange{630}{660}{\milli\kelvin} through primary CBT thermometry, which represent to our best knowledge the lowest verified in a cryogenic wafer prober. The on-wafer thermometry also revealed a systematic excess of approximately \qty{40}{\milli\kelvin} over a nominal chuck setpoint of \qty{600}{\milli\kelvin}. \textit{In situ} calibrated secondary CBT thermometry enabled resolving the thermal transients of each probe touchdown, confirming a thermalization within five minutes under our measurement conditions.

With the CWP verifiably operating below the critical temperature of aluminum, this approach could also enable direct wafer-scale screening of Al-based superconducting circuits and Josephson devices, including qubits \cite{anferovSuperconductingQubits202024}.

Our results establish that temperature-critical testing in a CWP cannot rely on chuck thermometry alone. An on-wafer temperature reference is required for accurate results. Our TiW CBTs are naturally suited to this role, either integrated on-chip with the devices under test or as a bespoke thermal reference wafer for commissioning, benchmarking, and monitoring cryogenic test platforms. The CBT secondary mode provides a measure of the on-wafer temperature in real time that could enable, for example, adaptive thermal settling schemes that replace fixed worst-case delays and can shorten overall testing time.

For the CBT devices, the dataset presented here also provides wafer-scale statistics of the TiW normal-metal junction process in an application-relevant cryogenic environment. The obtained distributions of charging energy and tunneling resistance reflect junction dimensions and tunnel barrier properties, parameters otherwise requiring destructive cross-sectional imaging or dedicated single-junction studies. Beyond the CBT parameters, additional information on the tunnel barrier can be extracted by fitting the high-bias background conductance with the Simmons model \cite{simmonsGeneralizedThermalJV1964}, yielding estimates of the effective barrier height and thickness. As the $E_\mathrm{C}$ extracted from the CBT conductance independently quantifies the junction capacitance, and, for a known junction area, the dielectric thickness, these two modalities provide a way to cross-validate the barrier parameters. Deviations from the ideal quadratic Simmons dependence (introduced in Eq. \ref{eq:Simmons}) can in turn be used to quantify barrier nonidealities, as addressed in Section \ref{sec:simmons}. Together, these methods offer a practical approach for process development and monitoring metrology using only DC conductance measurements.

Our CBT-CPW probing work shows that sub-kelvin reciprocal wafer-level testing is feasible and mutually beneficial for the instrument and the devices under test, providing the verified thermal environment and statistical processes required to carry wafer scale quantum hardware development from laboratory demonstrations towards industrial manufacturing.

\begin{acknowledgments}
We thank Nikolai Yurttagül for useful discussions on CBT modeling and physics, and for the Python libraries used as a basis for our analysis software. Alberto Ronzani is acknowleged for useful discussions on experimental methods.

This work was supported by the European Union's Horizon RIA and EIC programmes under Grant Nos. 824109, European Microkelvin Platform; 101113086, SoCool; and 101113983, Qu-Pilot. We also acknowledge support from the Research Council of Finland through Project No. 350667, Femto; Project No. 336817, QTF Centre of Excellence; and Project No. 374172, QMAT Centre of Excellence. Additional support was provided by Business Finland through the CryoTherm project and Project No. 128291, Quantum Technologies Industrial (QuTI), as well as by the Technology Industries of Finland Centennial Foundation and the Chips Joint Undertaking through Project No. 101139908, ARCTIC.

\end{acknowledgments}

\section{Data availability}

The data supporting the findings of this article are openly available in Zenodo at Ref. \cite{lehtisyrjaeCryogenicWaferProbingZenodo2026}.

\appendix
\section{\label{app:design}Device design and fabrication}

A high-quality junction process requires controlling the junction resistance and capacitance through careful definition of the junction dimensions and deposition of the tunnel barrier. This is key to achieving good device yield with reproducible performance and low thermometric uncertainty \cite{pekolaInfluenceDeviceNonuniformities2022}. Our junction fabrication is based on the sidewall passivated process (SWAPS) \cite{gronbergSidewallSpacerPassivated2017}, adapted to fabricate NIN tunnel junctions on wafer-scale. Here, the trilayer tunnel junctions comprise titanium tungsten (TiW) as an electrode material and \textit{in situ} oxidized aluminum (AlO$_\mathbf{x}$) as a tunnel barrier. Recently, we have demonstrated inherent normal state operation of these junctions down to \qty{20}{\milli\kelvin} without needing to apply external magnetic fields \cite{luomahaaraScalableNonsuperconductingTunnel2026}. Furthermore, the SWAPS process has previously proven performance, uniformity, and scalability in numerous devices based on superconducting tunnel junctions \cite{kivirantaTwoStageSQUIDAmplifier2021, hatinenEfficientElectronicCooling2024, perelshteinBroadbandContinuousVariableEntanglement2022}.

The target operating temperature range determines the required CBT charging energy, and consequently the junction capacitance and dimensions. The upper temperature limit of a given CBT is set by the resolution of the conductance measurement, specifically the minimum discernible conductance dip amplitude. Conversely, the lower temperature limit for operation in the universal regime is reached as $k_\mathrm{B}T$ approaches $E_\mathrm{C}$. Beyond this, in the intermediate Coulomb blockade regime, finite charging effects become significant, and random background charges begin to significantly influence the measured conductance \cite{giazottoOpportunitiesMesoscopicsThermometry2006, feshchenkoPrimaryThermometryIntermediate2013}. Thus, the practical dynamic range of a single CBT is roughly $10^2$ to $10^3$.

Practical CBT devices typically employ a two-dimensional array of tunnel junctions, with $N$ junctions connected in series in each row and $M$ rows connected in parallel. The series junctions broaden the conductance dip according to Eq.~\eqref{eq:v_half} and improve robustness against voltage noise, electromagnetic interference, electrostatic discharge, and offset voltages arising from, e.g. thermal electromotive force (EMF). The number of parallel rows sets the total array impedance, typically to approximately \qty{100}{\kilo\ohm} for voltage-biased measurements, and reduces sensitivity to local variations in junction resistance and random offset-charge distributions \cite{yurttagulCoulombBlockadeThermometry2021}. 

\section{\label{app:experiment}Experiment setup and procedure}
The Cryogenic Wafer Prober developed by Bluefors and AEM Afore employs \qty{50}{\kelvin} and \qty{4}{\kelvin} pulse tube stages and a final $^3$He cooling stage for sub-kelvin cooling. Flexible copper braids thermally couple the $^3$He stage to the wafer chuck and probe card holder, reducing temperature gradients between the wafer and probe tips (Fig.~\ref{fig:prober}). Chuck and probe card temperatures are monitored by resistance thermometers, read out by Bluefors Temperature Controllers \cite{EnhancedUserExperience}.

Vision for wafer alignment is provided by a microscope and windows through the outer vacuum chamber and radiation shields. The room temperature view port is equipped with a pneumatically actuated shutter, which was closed during measurements to minimize the radiation thermal load and any interference caused by ambient light.

Electrical contact to the DUTs on the wafer is achieved with a probe card having a total of 16 tungsten cantilever-style probes, arranged in two parallel staggered rows with a pitch of \qty{500}{\micro\meter}. 

The experiment wiring inside the cryostat to the probe card is realized with a Bluefors standard phosphor bronze twisted multipair loom \cite{blueforsoyTwistedPairWiring2021}, which contains a total of 24 lines with a wire gauge of 36 AWG, thermally anchored to each stage of the cryostat. The wires are passed through the cryostat outer vacuum chamber with a hermetic Fischer connector, to which a BNC breakout box was connected with a shielded twisted multipair cable. A cryogenic $\pi$-type low-pass filter with a corner frequency of \qty{100}{\kHz} was placed in-line, mounted on the 1 Kelvin flange, which also contains all probe card interfacing connectors.

The measurement setup follows the standard lock-in amplifier based differential conductance technique in the four-probe configuration, wherein the voltage signal is read out from the CBT devices via lines separate from the voltage biasing. In the differential conductance ($\frac{dI}{dV}$) measurement, a small AC excitation voltage is superimposed on the swept DC bias voltage, and the resulting AC current and voltage across the CBT are measured at each bias point.

\begin{figure}[h]
    \includegraphics[width=\linewidth]{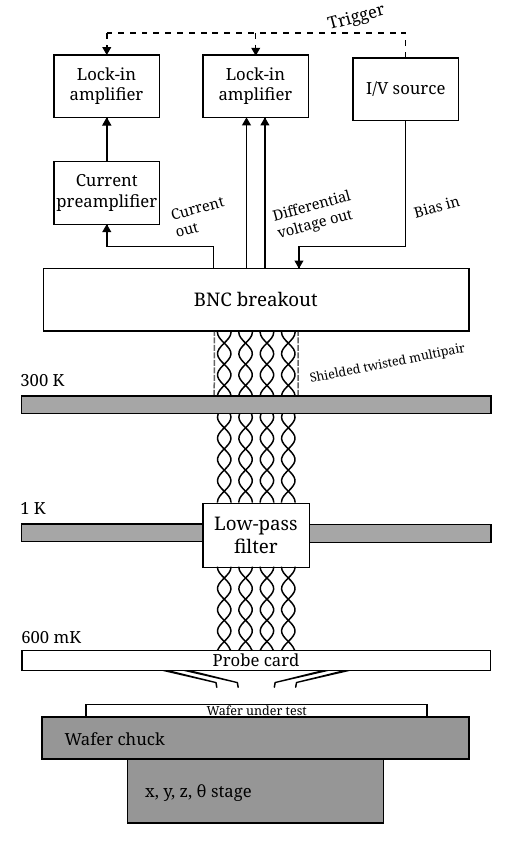}
    \caption{Diagram of the differential conductance measurement setup on the cryogenic wafer prober.}
    \label{fig:meas_setup}
\end{figure}

The combined AC and DC bias was generated using a Lake Shore 155 I/V source. The AC current response was measured using a Basel Precision Instruments SP983c current-to-voltage converter, with the output read by a Zurich Instruments MFLI lock-in amplifier. The AC voltage across the CBT was measured using a second MFLI unit. An AC peak-to-peak excitation voltage of \qty{1}{\mV} was used for all measurements, which is small enough not to "blur" the shape of the CBT dip at $\lesssim \qty{1}{\percent}$ of $V_{1/2}$ (Eq.~\eqref{eq:v_half}) at all temperatures, and large enough to provide a good signal-to-noise ratio for the lock-in measurement. The AC excitation frequency was set to \qty{63}{\Hz} for all measurements presented here, which was selected as the frequency that yielded the lowest total measurement noise, being sufficiently far from the $1/f$ noise knee and from the \qty{50}{\Hz} mains frequency and its harmonics. A diagram of the measurement setup is provided in Fig.~\ref{fig:meas_setup}.

Special attention was paid to minimizing capacitive noise pickup in the experiment wiring, owing to the relatively high impedance of the CBT devices. The cryostat is located in a factory production environment where electromagnetic interference (EMI) from, e.g., electric motors and their drives contaminates the mains supply and the surrounding space with conducted and radiated EMI. A well-shielded room-temperature multi-pair twisted cable and BNC breakout were employed, and proper shield continuity was verified throughout the wiring. The grounding scheme of the cryostat and measurement instruments was carefully optimized, with the lowest noise configuration achieved using a star grounding topology. Here, the cryostat was grounded separately via a large-gauge cable, and each instrument was grounded individually, with all ground connections terminating at the chassis of the BNC breakout box, which itself was connected to mains earth through a front panel connector on the Lake Shore 155. Of the 16 pins on the probe card, only 4 were used for the four-probe measurement, with the remaining pins left floating. To suppress noise pickup on these lines, BNC caps were fitted to each corresponding connector at the breakout box.

All measurement lines passing into the cryostat were filtered by a cryogenic $\pi$-type low-pass filter mounted on the \qty{1}{\kelvin} flange, with a corner frequency of \qty{100}{\kHz}. This corner frequency is sufficiently above the AC excitation frequency to leave the measurement signals unaffected, while attenuating high-frequency EMI conducted along the wiring from the room-temperature electronics or picked up from the environment. The filter introduces a small series resistance, which is inconsequential in the four-probe configuration.

The wafer probing procedure was as follows. The wafer was manually loaded onto the chuck, roughly aligned by hand, and secured with three edge clamps. The radiation shields were then assembled and the outer vacuum chamber closed. The rotational alignment of the wafer was carried out using the rotational stage of the chuck, after which the intra-reticle DUT coordinates were taught, and the initial contact height was established using a dedicated edge-sensor probe, configured to detect contact with the wafer surface. The edge sensor was disabled during further measurements, as its circuit introduced noise to the measurement. The wafer map, reticle layout, and DUT coordinates were defined in the Afore prober control software Hermes. Still at room temperature, automatic stepping to each DUT was tested, and galvanic contact of the probe needles to the samples was verified by two-wire resistance measurement with a Keysight 34465A digital multimeter. The system was then prepared for cooldown by evacuating the outer vacuum chamber and initiating the pulse tubes. The final cooldown to base temperature was achieved by starting the $^3$He circulation. Once the system reached base temperature, wafer navigation and contact heights were verified, and any drift caused by differential thermal contraction was compensated.

The quality of probe contact was found to vary across the DUTs and temperature points, with some measurements exhibiting excess noise from poor contact and consequently being excluded from the analysis. Achieving reliable probe contact at cryogenic temperatures is a recognized challenge. The contact resistance of probe–pad interfaces known to be low at room temperature cannot in general be assumed to remain low at cryogenic temperatures. The mechanisms governing this behavior were not characterized as part of the present work and are not documented in the literature. A plausible contributing factor, supported by adjacent literature, is the low temperature hardening of the contact pad material \cite{iwabuchiDevelopmentVickerstypeHardness1996}, which may degrade the contact resistance of the probe-pad interface \cite{naglerImprovedModelElectrical2019}.

The chuck temperature was varied between the base temperature of the cryogenic prober ($\approx$ \qty{600}{\milli\kelvin}) and \qty{1.95}{\kelvin}, the highest temperature achievable without overloading the $^3$He circulation. PID temperature control, provided by the Bluefors Temperature Controllers, was used to stabilize the chuck at each setpoint. 

The validity of the primary and the secondary CBT thermometry as a real-time thermalization monitor rests on the assumption that the electron temperature in the CBT islands closely follows the local phonon bath temperature. At the temperatures accessed in this work, the electron–phonon heat flux is sufficiently strong that thermalization of the electron system to the phonon bath is rapid compared to the measurement timescale. Electron temperature saturation associated with weak electron-phonon coupling has been observed only below approximately \qty{20}{\milli\kelvin} in nearly identical devices \cite{luomahaaraScalableNonsuperconductingTunnel2026} and in comparable CBTs \cite{meschkeElectronThermalizationMetallic2004}. This is well below the temperature range investigated here. Heat load from Joule heating by the AC and DC excitation voltages is estimated to be not significant at these temperatures. Electron overheating can therefore be disregarded in the present measurement conditions, and the CBT electron temperature is expected to accurately reflect the local phonon bath temperature of the wafer at all measurement points.

\section{\label{app:data}Complete dataset}
\begin{figure*}[p!]
    \includegraphics[width=\linewidth]{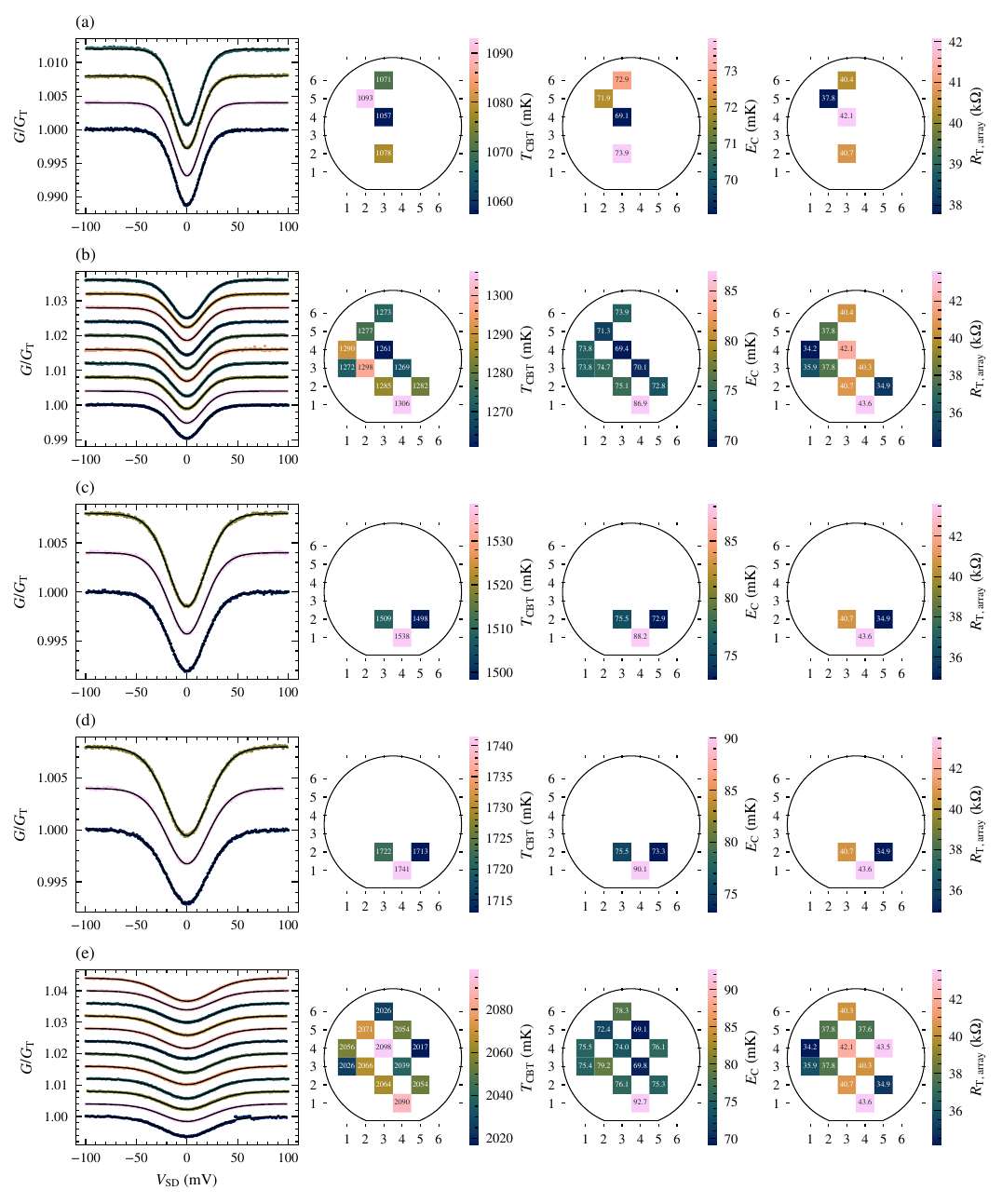}
    \caption{Compensated conductance data and extracted CBT parameters as wafer maps for chuck temperatures $T_\mathrm{chuck}$ (a) \qty{1.05}{\kelvin}, (b) \qty{1.25}{\kelvin}, (c) \qty{1.45}{\kelvin}, (d) \qty{1.65}{\kelvin}, and (e) \qty{1.95}{\kelvin}.}
    \label{fig:fulldata}
\end{figure*}

The complete set of background-compensated conductance data remaining after the exclusion of noisy measurements, together with the parameters extracted by fitting the composite model at each chuck-temperature setpoint, is presented in Fig.~\ref{fig:fulldata}. The \qty{600}{\milli\kelvin} data are omitted because they are already presented in Fig.~\ref{fig:G_and_wafermaps}.

\clearpage
\bibliography{cbt_cwp}

\end{document}